\documentclass[prd,aps,showpacs,eqsecnum,floatfix,nofootinbib,preprint,tightenlines]{revtex4}

\usepackage{latexsym}
\usepackage{epsf}
\usepackage{graphicx}
\usepackage{slashed}
\usepackage{amsmath}
\usepackage{amssymb}
\usepackage{multirow}
\def\p{\pi}                      % Also, \varpi
\def\cbo{{\,\raise-.15ex\Sc [\,}}                       % curly "
\def\ddt#1{{\buildrel {\hbox{\LARGE .\kern-2pt.}} \over {#1}}}% double dot-over
\newcommand{\gA}{g_A}
\newcommand{\gAb}{\bar{g}_A}
\newcommand{\OV}[1]{V^{a}_{{#1}}}
\newcommand{\OA}[1]{A^a_{{#1}}}
\newcommand{\OT}[1]{T^a_{{#1}}}

\long\def\symbolfootnote[#1]#2{\begingroup%
\def\thefootnote{\fnsymbol{footnote}}\footnote[#1]{#2}\endgroup}

\long \def \blockcomment #1\endcomment{}

\def\vp{{\vec p}}

\def\vn{{\vec n}}

\newcommand{\pref}[1]{(\ref{#1})}

\newcommand{\mN}{M_N}

\newcommand{\mf}{\langle x\rangle_{u-d}}
\newcommand{\hm}{\langle x \rangle_{\Delta u - \Delta d}}
\newcommand{\tm}{\langle x \rangle_{\delta u - \delta d}}
\newcommand{\Et}{E_{N\pi}}
\newcommand{\av}{a^v_{2,0}}
\newcommand{\dav}{\Delta a^v_{2,0}}

\begin{document}
\hyphenation{fer-mio-nic per-tur-ba-tive pa-ra-me-tri-za-tion
pa-ra-me-tri-zed a-nom-al-ous}

\renewcommand{\thefootnote}{$*$}

\preprint{YITP-16-143}

\title{Nucleon-pion-state contribution in lattice calculations of moments of parton distribution functions}

\author{Oliver B\"ar$^{a}$} 
\affiliation{$^a$Yukawa Institute for Theoretical Physics, Kyoto University,
\\Kitashirakawa Oiwakechou, Sakyo-ku, Kyoto 606-8502, Japan\\
}

\begin{abstract}
\vspace{0.5cm}

We employ chiral perturbation theory to calculate the nucleon-pion-state contribution to the 3-point correlation functions measured in lattice QCD to compute various moments of parton distribution functions (quark momentum fraction, helicity and transversity moment).  We estimate the impact of the nucleon-pion-state contribution on the plateau method for lattice simulations with a physical pion mass. The nucleon-pion-state contribution results in an overestimation of all three moments. The overestimation is at the 5-20\% level for source-sink separations of about 1.5 fm. 

\end{abstract}

\maketitle

\renewcommand{\thefootnote}{\arabic{footnote}} \setcounter{footnote}{0}

\newpage

\section{Introduction}\label{Intro}

Lattice QCD calculations of hadron structure observables have been actively pursued for a long time. Despite continuous progress many lattice results still show sizeable deviations from the experimentally measured values. A prominent example is the isovector quark momentum fraction $\mf$. Compilations of the lattice efforts to compute this observable can be found in various recent reviews \cite{Syritsyn:2014saa,Green:2014vxa,Alexandrou:2016hiy}, but it seems fair to summarize them by saying that essentially all lattice calculations overestimate the quark momentum fraction by 30-60\%.

Lattice QCD simulations are afflicted with various systematic uncertainties. For light quark masses larger than their physical values a chiral extrapolation of the lattice data to the physical point has to be made. Results of Chiral Perturbation Theory (ChPT) are usually employed for this step, but a large chiral extrapolation is considered to be problematic. How well the chiral extrapolation is behaved depends on the physical quantity, and $\mf$ seems to be particularly sensitive in that respect.

Progress in computer power and simulation algorithms in the past few years have made lattice simulations possible with physical light quark masses. Such `physical point simulations' require no chiral extrapolation, thus eliminating a major source of uncertainty. Recently, the Regensburg QCD (RQCD) collaboration and the European twisted mass (ETM) collaboration have reported results for $\mf$ obtained in physical point simulations \cite{Bali:2014gha,Abdel-Rehim:2015owa}.\footnote{For simplicity we refer to lattice simulations with pion masses smaller than about 150 GeV as physical point simulations. The small mismatch to the physical point is irrelevant for our discussion.} The lattice results still deviate by about 25\% from the experimental value.

Another source of uncertainty are excited-state contaminations in the correlation functions measured on the lattice to calculate the hadronic observables. In fact, these contaminations become more severe the smaller the quark masses are. In physical point simulations one expects two-particle nucleon-pion ($N\pi$) states to contribute substantially to the excited-state contamination in hadronic correlation functions. The small physical pion mass implies that the energy of a $N\pi$ state can be smaller than the energy of the first resonance state, provided the discrete and opposite spatial momenta of the nucleon and pion are sufficiently small. For typical lattice volumes with $M_{\pi}L\approx4$ and periodic boundary conditions this is the case for three $N\pi$ states. For larger volumes satisfying $M_{\pi}L\approx6$, as realized in the simulations of the PACS collaboration \cite{Ishikawa:2015rho}, this number increases to six. This raises the concern whether the advantage of physical point simulations is compromised by stronger excited-state contaminations.  

In this paper we follow up on a recent ChPT calculation \cite{Bar:2016uoj} of the $N\pi$-state contribution in lattice determinations of the nucleon axial, tensor and scalar charge. That ChPT can be employed to compute multiparticle-state contributions involving light pions has  been proposed already some time ago \cite{Tiburzi:2009zp,Bar:2012ce}. Here we compute the $N\pi$-state contribution to three Mellin moments of parton distribution functions (PDFs): the quark momentum fraction $\mf$, the helicity moment $\hm$ and the transversity moment $\tm$. At leading twist, these moments can be extracted from nucleon matrix elements involving local one-derivative vector, axial-vector and tensor operators. As non-singlet quantities their lattice calculation does not involve disconnected contributions, thus these moments are, together with the nucleon axial, tensor and scalar charge, the simplest hadron structure observables one can measure on the lattice. Recently, the ETM collaboration has presented results for all six observables obtained in physical point simulations \cite{Abdel-Rehim:2015owa}.

The computation presented here parallels the one for the nucleon charges. We employ the covariant formulation of Baryon ChPT \cite{Gasser:1987rb,Becher:1999he} to leading order (LO) in the chiral expansion. At this order the low-energy coefficients (LECs) entering the results are all known from phenomenology. Thus, we obtain definite predictions for the $N\pi$-state contribution to the three moments estimated by the plateau method. On the other hand, as long as the next-to-leading order (NLO) corrections are not known it is difficult to assess the error of the LO results. Still, crude estimates can be made and we obtain, for example, a 10-20\% overestimation in case of $\mf$ due to the $N\pi$ contribution using the plateau method at source-sink separations of about 1.5 fm. Even though not very precise this number indicates that the $N\pi$-state contamination may be responsible for a substantial part of the discrepancy still observed between the lattice results and the experimental value.

\section{Moments of parton distribution functions}
\subsection{Basic definitions}
Throughout this paper we consider QCD with the simplification of equal up and down quark masses. We work in euclidean space-time with infinite time-extent. The spatial volume, however, is taken to be finite with extent $L$ in each spatial direction, and periodic boundary conditions are imposed. 

We will be interested in the forward nucleon matrix elements $\langle N(p)|{\cal O}_X|N(p)\rangle$, where the operator ${\cal O}_X$ with $X=V,A,T$ stands for one of the following local one-derivative operators\footnote{We follow to a large extent the conventions in Ref.\ \cite{Abdel-Rehim:2015owa}.}
\begin{eqnarray}
\OV{\mu\nu} &=& \overline{q}\gamma_{{\{\mu}}D^-_{\nu\}} T^a q\,,\label{vectop}\\
\OA{\mu\nu} &=& \overline{q}\gamma_{{\{\mu}}D^-_{\nu\}}\gamma_5 T^a q\,,\label{axvectop}\\
\OT{{\mu\nu\rho}} &=& \overline{q}\sigma_{{[\mu\{\nu]}}D^-_{\rho\}} T^a q\,.\label{tensorop}
\end{eqnarray}
$q=(u,d)^T$ denotes the isospin quark doublet and the (color covariant) derivative is defined as
\begin{equation}\label{symDer}
D^-_{\mu} = \frac{1}{2} \Big( {\overrightarrow D}_{\mu} -  \overleftarrow{D}_{\mu} \Big)\,.
\end{equation}
The curly and square brackets refer to symmetrization and antisymmetrization, respectively. Symmetrization also involves subtracting the trace. The SU(2) generators are defined as half of the Pauli matrices, $T^a=\sigma^a/2$.

From the forward matrix elements of these operators one can obtain the first three moments of the PDFs, the momentum fraction $\mf$, the helicity moment $\hm$ and the transversity moment $\tm$. This is conveniently done by computing the ratio
\begin{eqnarray}
R_{X}(\Gamma_{\nu},t,t')=\frac{G_{{\rm 3pt},X}(\Gamma_{\nu},t,t')}{G_{\rm 2pt}(t)}\,.
\end{eqnarray}
of the 3-point (pt) and 2-pt functions
\begin{equation}\label{Def3pt}
G_{{\rm 3pt},X}(\Gamma_{\nu},t,t') = \int d^3x\int d^3y \,\Gamma'_{\nu,\alpha\beta} \langle N_{\beta}(\vec{x},t) {\cal O}_X^3(\vec{y},t') \overline{N}_{\alpha}(\vec{0},0)\rangle\,,
\end{equation}
\begin{equation}\label{Def2pt}
G_{\rm 2pt}(t) = \int d^3x \,\Gamma_{4,\alpha\beta} \langle N_{\beta}(\vec{x},t) \overline{N}_{\alpha}(\vec{0},0)\rangle\,.
\end{equation}
Here $N, \overline{N}$ are interpolating fields for the nucleon. For the projection matrices $\Gamma_{\nu}$ we follow Ref.\ \cite{Abdel-Rehim:2015owa} and define
\begin{equation}\label{DefGammaNu}
\Gamma_4 = \frac{1}{4} (1+\gamma_4), \quad \Gamma_{k}  = \Gamma_4 i \gamma_5 \gamma_k\,.
\end{equation} 
Performing the standard spectral decomposition of the two correlation functions and taking all times $t,t'$ and $t-t'$ to infinity it is straightforward to show that the ratio $R_X$ goes to a constant,
\begin{equation}\label{RAsympt}
R_X(\Gamma_{\nu},t,t') \longrightarrow \Pi_X(\Gamma_{\nu})\,.
\end{equation}
According to our definitions this constant is related to the various moments in the following way \cite{Abdel-Rehim:2015owa}:
\begin{eqnarray}
\Pi_{V_{44}}(\Gamma_4)& =& -\frac{3M_N}{4} \mf\,,\nonumber\\
\Pi_{A_{j4}}(\Gamma_k) &= &-\frac{i}{2}\delta_{jk} M_N \hm \,,\label{PiValues}\\
\Pi_{T_{\mu\nu\rho}}(\Gamma_k) &=& i\epsilon_{\mu\nu\rho k}\frac{M_N}{8} (2\delta_{4\rho} -\delta_{4\mu}-\delta_{4\nu}) \tm\,.\nonumber
\end{eqnarray}
For finite euclidean times $t,t'$ the ratio contains corrections which are exponentially suppressed. These stem from resonances and multihadron states that have the same quantum numbers as the nucleon. For small pion masses the dominant multihadron states are two-particle $N\pi$ states with the nucleon and the pion having opposite spatial momenta. Taking only these corrections into account the asymptotic behavior of the ratio reads
\begin{eqnarray}
\label{RatioAsymp}
R_X(\Gamma_{\nu},t,t') =  \Pi_X(\Gamma_{\nu})\Big[1+ \sum_{\vec{p}_n} \left(b_{X,n} e^{-\Delta E_n (t-t')} + \tilde{b}_{X,n} e^{-\Delta E_n t'} + \tilde{c}_{X,n} e^{-\Delta E_n t }\right)\Big].
\end{eqnarray}
Since we assume a finite spatial volume the momenta are discrete and the sum runs over all momenta allowed by the boundary conditions we impose. $\Delta E_n = E_{N\pi,n} -M_N$ is the energy gap between the nucleon-pion state and the ground state describing a nucleon at rest. Because the pions interact weakly with the nucleons the total energy $E_{N\pi,n}$ equals approximately the sum $E_{N,n}+E_{\pi,n}$ of the nucleon and pion energy, $E_{N,n}=\sqrt{p_n^2 +M_N^2}$ and $E_{\pi,n}=\sqrt{p_n^2 +M_{\pi}^2}$.
The coefficients $b_{X,n},\tilde{b}_{X,n}$ and $\tilde{c}_{X,n}$ in \pref{RatioAsymp} are dimensionless ratios of various matrix elements involving the nucleon interpolating fields and the operator ${\cal O}_X$. For example, the coefficient $\tilde{c}_{X,n}$ contains the excited-to-excited-state matrix element $\langle N(\vec{p}_n)\pi(-\vec{p}_n)| {\cal O}_X|N(\vec{p}_n)\pi(-\vec{p}_n)\rangle$.\footnote{Similar contributions involving this matrix element with different momenta in the initial and final state will be ignored in the following.} 

\subsection{The chiral effective theory}\label{chpt}

The correlation functions and the ratios $R_X$ defined in the previous section can be computed in the chiral effective theory of QCD, i.e.\ in ChPT. For sufficiently large times $t,t'$ pion physics will dominate the correlation functions and ChPT is expected to provide good estimates for them. Similar calculations for the 3-pt functions involving the axial vector current as well as the tensor and scalar density have been performed in Ref.\ \cite{Bar:2016uoj}, and the result for the 2-pt function can be found in Ref. \cite{Bar:2015zwa}. The calculation presented here is analogous to the ones in these two references, the main difference is the different set of operators entering the 3-pt functions. In order to compute them in the chiral effective theory we need the ChPT expressions for the three operators in \pref{vectop} - \pref{tensorop}.
  
Our calculations are performed in the covariant formulation of baryon ChPT \cite{Gasser:1987rb,Becher:1999he}. 
Based on the transformation properties under chiral symmetry, parity and charge conjugation the ChPT expressions for the  operators \pref{vectop}, \pref{axvectop} have been constructed in \cite{Dorati:2007bk,Wein:2014wma}. Since we work to LO we only need the leading contributions. In terms of the nucleon fields $\Psi=(p,n)^T$ and $\overline{\Psi}=(\overline{p},\overline{n})$, which contain the Dirac fields for the proton $p$ and the neutron $n$, we find\footnote{We follow the notation in Ref.\ \cite{Dorati:2007bk}. The expressions \pref{OpVector}, \pref{OpAxialVector} are easily obtained from the source term given in eq.\ (21) in that reference.}
 \begin{eqnarray}
\OV{\mu\nu} &=& a_{2,0}^v \overline{\Psi} \gamma_{\{\mu}\partial_{\nu\}}^- \sigma^a\Psi -  \frac{\Delta a_{2,0}^v}{f} \epsilon^{abc} \pi^b \overline{\Psi} \gamma_{\{\mu}\gamma_5\partial_{\nu\}}^- \sigma^c\Psi\,,\label{OpVector}\\
\OA{\mu\nu} &=& \Delta a_{2,0}^v \hat{S} \overline{\Psi} \gamma_{\{\mu}\gamma_5\partial_{\nu\}}^- \sigma^a\Psi -  \frac{a_{2,0}^v}{f} \epsilon^{abc} \pi^b \overline{\Psi} \gamma_{\{\mu}\partial_{\nu\}}^- \sigma^c\Psi\,.\label{OpAxialVector}
\end{eqnarray}
Here we have already expanded in powers of pion fields up to linear order, since this is sufficient for our calculation. The derivative $\partial_{\mu}^{-} = ({\overrightarrow \partial}_{\mu} - \overleftarrow{\partial}_{\mu})/2$ contains the standard partial derivatives acting on the nucleon fields. Besides the LO LEC  $f$, the pion decay constant in the chiral limit, these expressions also contain two more LECs, $a_{2,0}^v$ and $\Delta a_{2,0}^v$. There normalization was chosen such that they correspond to the chiral limit values of the momentum fraction $\mf$ and the helicity moment $\hm$, respectively.

The  results in \pref{OpVector}, \pref{OpAxialVector} resemble the expressions for the vector and axial vector currents. There are two contributions and their LECs are related due to chiral symmetry. On the other hand, there is no contribution involving only pion fields. The reason is that Lorentz indices in terms with pion fields can only come from partial derivatives, and we need at least two of those to form a symmetric tensor. Such an expression is necessarily two orders higher in the chiral counting, as has been discussed in Ref.\ \cite{Dorati:2007bk}. 

The expression for the tensor operator \pref{tensorop} in covariant ChPT has, to our knowledge, not been constructed yet. Since we need the LO expression only the construction is straightforward. We defer the details of the construction to appendix \ref{appTensor}, here we just quote the final result. To leading chiral dimension we find only one term, 
\begin{equation}
\OT{\mu\nu\rho} = \delta a_{2,0}^v \overline{\Psi} \sigma_{[\mu \{\nu ]}\partial_{\rho\}}^- \sigma^a\Psi \,.\label{OpTensor}
\end{equation}
Here too we have already expanded in powers of pion fields, and we dropped all contributions involving two or more of them. The LEC $\delta a_{2,0}^v$ associated with this term is chosen such that it corresponds to the chiral limit value of the transversity moment $\tm$. Also for the tensor operator there is no purely pionic contribution at leading chiral dimension.

For the calculation of the correlation functions we also need the Feynman rules stemming from the chiral Lagrangian and the nucleon interpolating fields. These are the same expressions as in Refs.\ \cite{Bar:2016uoj,Bar:2015zwa}. For completeness and for the readers convenience we summarize them in Appendix \ref{appFeynmanRules}, and refer to \cite{Bar:2016uoj,Bar:2015zwa} for more details concerning their derivation.

\subsection{The 3-pt functions in ChPT}
\label{ssect3ptfunctions}
% Figure
% Diagram for single nucleon contribution
\begin{figure}[tp]
\begin{center}
\includegraphics[scale=0.5]{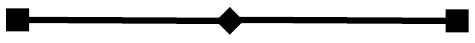}\\[3ex]
\caption{Leading Feynman diagram for the 3-pt function. Squares denote the nucleon interpolating fields at times $t$ and $0$, the diamond represents  the operator at insertion time $t'$. Solid lines stand for the nucleon propagators. The two integrations in \pref{Def3pt} imply zero spatial momentum propagators in this diagram.}
\label{fig:diagNOX}
\end{center}
\end{figure}
% End figure
The perturbative calculation of the 3-pt functions in ChPT is straightforward. The computation parallels the one in Ref.\ \cite{Bar:2016uoj} where the 3-pt functions involving the axial-vector current, the tensor and the scalar density are computed. The calculation is conveniently done using the time-momentum representation of the finite volume propagators for the nucleon and pion (see appendix \ref{appFeynmanRules}).

The leading diagram for the 3-pt functions is shown in fig.\ \ref{fig:diagNOX}. It gives the leading single-nucleon-state contribution $G^{N}_{{\rm 3pt},X}$ to the 3-pt function, and the result reads
\begin{equation}\label{SingleNucl}
G^{N}_{{\rm 3pt},X}(\Gamma_{\nu},t,t') = \Pi_X(\Gamma_{\nu}) G^{N}_{{\rm 2pt}}\,,
\end{equation}   
with the constant $\Pi_X$ defined in \pref{PiValues}. $G^{N}_{{\rm 2pt}}$ denotes the leading single-nucleon-state contribution to the 2-pt function \cite{Bar:2015zwa}.\footnote{The definition for $\Gamma_4$ in Ref.\ \cite{Bar:2015zwa} differs by a factor 2 from the one in \pref{DefGammaNu}, such that $G^{N}_{{\rm 2pt}}$ needs to be divided by 2.} Forming the ratio of the 3-pt and 2-pt function we find $R_X=\Pi_X$ in accordance with \pref{RAsympt}. 

\begin{figure}[t]
%\begin{center}
\includegraphics[scale=0.45]{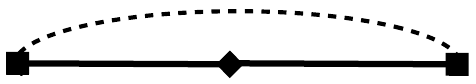}\hspace{0.3cm}\includegraphics[scale=0.45]{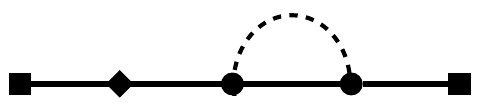}\hspace{0.3cm}\includegraphics[scale=0.45]{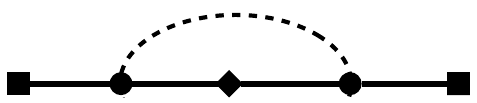}\hspace{0.3cm}\includegraphics[scale=0.45]{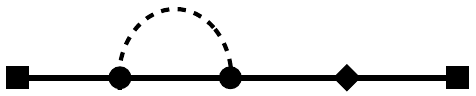}\\
a)\hspace{3.5cm} b)\hspace{3.5cm} c)\hspace{3.5cm} d)\\[3ex]
\includegraphics[scale=0.45]{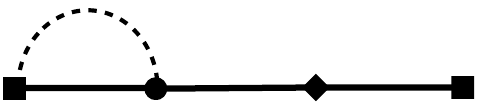}\hspace{0.3cm}\includegraphics[scale=0.45]{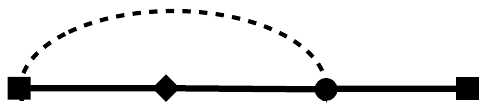}\hspace{0.3cm}\includegraphics[scale=0.45]{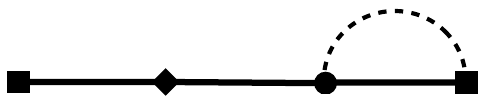}\hspace{0.3cm}\includegraphics[scale=0.45]{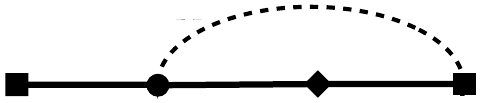}\\
e)\hspace{3.5cm} f)\hspace{3.5cm} g)\hspace{3.5cm} h) \\[3ex]
\includegraphics[scale=0.45]{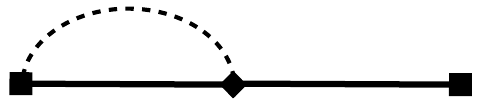}\hspace{0.3cm}\includegraphics[scale=0.45]{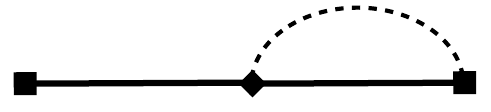}\hspace{0.3cm}\includegraphics[scale=0.45]{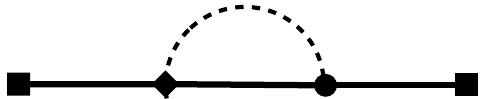}\hspace{0.3cm}\includegraphics[scale=0.45]{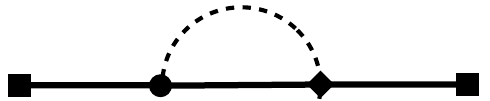}\\
i)\hspace{3.5cm} j)\hspace{3.5cm} k)\hspace{3.5cm} l)\\[0.3ex]
\caption{Feynman diagrams for the LO nucleon-pion contribution in the 3-pt functions. Circles represent a vertex insertion at an intermediate space time point, and an integration over this point is implicitly assumed. The dashed lines represent pion propagators.}
\label{fig:VCNpidiags}
\end{figure}

Figure \ref{fig:VCNpidiags} displays the diagrams with a nonzero $N\pi$-state contribution to the 3-pt functions. Diagrams a) - h) contribute to all three correlation function ($X=V,A,T$). Diagrams i) - l) contribute to $X=V,A$ only because the tensor operator does not contain a $\overline{\Psi}\Psi\pi$ term.
It is convenient to express the $N\pi$-state contribution $G^{N\pi}_{{\rm 3pt},X}$ to the 3-pt function in the form (for notational simplicity we drop the subscript $n$ on the coefficients in this section)
\begin{equation}\label{NuclPionContrGeneral}
G^{N\pi}_{{\rm 3pt},X}= G^{N}_{{\rm 3pt},X} \sum_{\vec{p}_n}\left(b_{X} e^{-\Delta E_n (t-t')} + \tilde{b}_{X} e^{-\Delta E_n t'}  + c_{X} e^{-\Delta E_n t} \right).
\end{equation}
% Figure
% Diagrams for nucleon pion contributions
% End figure

In order to quote our results for the coefficients we introduce the same short hand notation as in Ref.\ \cite{Bar:2016uoj}. 
We introduce ``reduced coefficients" $B_X, C_X$ that differ from the original ones by two overall factors that are common to all coefficients: 
\begin{eqnarray}
b_{X} &=& \frac{1}{16(fL)^2E_{\pi}L} \left(1-\frac{M_N}{E_N}\right)B_{X}\,,\label{finalb}\\
c_{X} &=& \frac{1}{16(fL)^2E_{\pi}L} \left(1-\frac{M_N}{E_N}\right)C_{X}\,.\label{finalc}
\end{eqnarray}
As in the calculation of the nucleon charges we explicitly find 
\begin{equation}\label{eqcoeff}
\tilde{b}_{X}=b_{X}
\end{equation}
for all three 3-pt functions, so we quote $b_{X}$ only. The first factor in \pref{finalb}, \pref{finalc} displays the expected $1/L^3$ dependence of a two-particle state in a finite volume. The second factor vanishes if the nucleon is, together with the pion, at rest. This has to be the case since the state with both nucleon and pion at rest is parity-odd, thus it cannot contribute to the 3-pt functions with parity-even nucleon interpolating fields.

The non-trivial results of our ChPT calculation are the remaining coefficients $B_X, C_X$. For them we find\footnote{In case of the axial-vector and tensor operators we show the results for the averaged correlation functions, where the average is taken over the indices $j$ in case of $A^a_{j4}$ and $k,\mu,\nu$ for the tensor operator $T^a_{\mu\nu4}$. The coefficients for the averaged correlation functions are slightly simpler than those for fixed indices.  However, the final results for the $N\pi$-state contribution are the same in both cases since the sum over the spatial momenta in \pref{RatioAsymp} also averages over the spatial directions.}
\begin{eqnarray}
C_{V} &=& -\frac{1}{3}\left(\gAb -1\right)^2 \left(4\frac{E_N}{M_N}-\frac{M_N}{E_N}\right)\,, \\ 
C_{A} &=& -\frac{1}{3}\left(\gAb -1\right)^2 \left(2\frac{E_N}{M_N} - 2 -\frac{M_N}{E_N}\right)\,,\\
C_{T} &=& +\frac{1}{3}\left(\gAb -1\right)^2 \left(3\frac{E_N}{M_N}-1 -\frac{M_N}{E_N}\right)\,,
\end{eqnarray}
where we have introduced the combination 
\begin{equation}
\gAb=g_A\frac{E_{N\pi} + M_N}{E_{N\pi} -M_N}\,, 
\end{equation}
with $E_{N\pi}=E_N+E_{\pi}$, $E_N=\sqrt{p^2+M_N^2}$ and $E_{\pi}=\sqrt{p^2+M_{\pi}^2}$.
For the coefficients $B_X$ we write $
B_X= \left(\gAb -1\right) \bar{B}_X$
with 
\begin{eqnarray}
\bar{B}_{V} &=& \frac{4}{3}\gAb \left[\frac{E_N}{M_N}+2\right]+g_A\left[\frac{\Et}{M_N}+1 \right] -\frac{8}{3}\frac{\dav}{\av}\left[\frac{E_N}{M_N}+1\right] -\Delta_B ,\label{BV}\\
\bar{B}_{A} &=& \frac{2}{3}\gAb \left[\frac{E_N}{M_N}+3\right]+g_A\left[\frac{\Et}{M_N}+\frac{5}{3} + \frac{2}{3}\frac{E_N}{M_N}\right] -\frac{8}{3}\frac{\av}{\dav}\left[\frac{E_N}{M_N}+1\right]-\Delta_B,\\
\bar{B}_{T} &=& \gAb \left[\frac{11}{3} - \frac{E_N}{M_N}\right]+g_A\left[2\frac{\Et}{M_N}+\frac{5}{3}- \frac{1}{3}\frac{E_N}{M_N} \right]+ \Delta_B \,.
\end{eqnarray}
and 
\begin{equation}
\Delta_B = \frac{2}{3}g_A\frac{M_{\pi}^2}{2E_{\pi}M_N-M_{\pi}^2} \,.
\end{equation}
Forming the ratio $R_X$ of the 3-pt and 2-pt functions we obtain expression \pref{RatioAsymp} with the coefficients
\begin{eqnarray}
\tilde{c}_{X} = c_X - c_{\rm 2pt}\,.
\end{eqnarray}
The coefficient $c_{\rm 2pt}$ entering the 2-pt function is given in \cite{Bar:2015zwa} and reads
\begin{eqnarray}
c_{\rm 2pt} &=& \frac{1}{16(fL)^2E_{\pi}L} \left(1-\frac{M_N}{E_N}\right)C_{\rm 2pt}\,,\quad C_{\rm 2pt} \,=\, 3\left(\gAb -1\right)^2\,.
\end{eqnarray}
The coefficients $b_X,\tilde{c}_X$ do not depend on the LECs associated with the interpolating nucleon fields; these cancel in the ratio. Thus, the LO results we have derived here are universal and apply to pointlike and smeared interpolating fields. This universality property, however, will be lost at the next order in the chiral expansion.
 
The coefficients $b_T,\tilde{c}_T$ for the tensor operator depend on two LECs only, $f$ and $g_A$, which are known experimentally very well. The coefficients for the vector and axial operator depend also on the ratio $\av/\dav$, i.e.\ on the ratio of the chiral limit values for $\mf$ and $\hm$. This ratio can be inferred from the experimentally measured values of the momentum fraction and the helicity moment. Therefore, the LO results derived here provide definite predictions for the $N\pi$ contributions to the ratios $R_X$, as discussed in the next section. 

The results for the coefficients simplify significantly in the heavy baryon (HB) limit that is obtained by sending the nucleon mass to infinity.
If we expand $E_N\approx M_N +p^2/2M_N$ in \pref{finalb}, \pref{BV} and drop all but the leading terms we obtain
\begin{equation}
b^{HB}_V= \frac{g_A^2}{2(fL)^2(E_{\pi}L)} \frac{p^2}{E_{\pi}^2}
\end{equation}
as the HB limit of the coefficient $b_V$. Similarly, we obtain for the remaining coefficients the results
\begin{equation}\label{HBlimitValues}
b^{HB}_A=b^{HB}_T=\frac{2}{3}b_V^{HB}\,,\qquad \tilde{c}_X^{HB} = - b_X^{HB}\,.
\end{equation}
Note that the HB limit values stem from the terms proportional to $\gAb^2$, all other terms vanish in the limit of infinite nucleon mass. In particular, the terms proportional to the ratio $\av/\dav$ are subleading and do not enter $b^{HB}_V,b^{HB}_A$.
From \pref{HBlimitValues} we would conclude that the $N\pi$-state contributions are equal for the axial and tensor operator, but fifty percent larger for the vector operator. In the next section we will see that this simple conclusion is modified once we are away from the HB limit.

The HB limits of the coefficients are also easily compared to their counterparts associated with the nucleon axial, scalar and tensor charges derived in \cite{Bar:2016uoj}. It turns out that the coefficients for the vector operator are equal to those associated with the scalar nucleon charge, i.e.\ $b^{HB}_V=b^{HB}_S$ and $\tilde{c}^{HB}_V=\tilde{c}^{HB}_S$. Similarly, the coefficients for the remaining two operators are equal to their analogues for the nucleon axial and tensor charge. This is in accordance with the expectation that the $N\pi$ contributions to the three moments and the three nucleon charges should be of the same order of magnitude, basically because the operators for all six observables are very similar in LO ChPT. 

\section{Impact on lattice calculations of the moments}

\subsection{Preliminaries}

In this section we estimate the impact of the $N\pi$-state contribution on the determination of the moments in lattice simulations. In order to do this we first need to fix the various input parameters that enter our results. 

The final result for the ratio $R_X$ can be written as
\begin{eqnarray}\label{simplnotationR}
R_X(t,t') &=& \Pi_X\Big[1+ \sum_{n\le n_{\rm max}} b_{X,n}\left(e^{-\Delta E_n(t-t')} + e^{-\Delta E_n t'}\right) + \tilde{c}_{X,n} e^{-\Delta E_n t} \Big]
\end{eqnarray}
if we make use of $b_{X,n}= \tilde{b}_{X,n}$, cf.\  eq.\ \pref{eqcoeff}. For notational simplicity we suppress the dependency on $\Gamma_{\mu}$ when writing $R_X$ and $\Pi_X$ in the following.  The coefficients $b_{X,n},\tilde{c}_{X,n}$ are dimensionless and depend on five independent dimensionless parameters: $\gA, f/M_N, M_{\pi}/M_N$, $M_{\pi}L$ and, in case of the vector and axial-vector operators, on the ratio $\av/\dav$.  To LO  we can use the experimental values for the LECs, i.e.\ we set  $\gA=1.27$, $f=f_{\pi}= 93$ MeV. The ratio $\av/\dav$ is approximately given by $\mf/\hm=0.165/0.19$ \cite{Alekhin:2012ig,Blumlein:2010rn}. We ignore the errors in the experimental values since they are too small to be significant in the following.

We are mainly interested in $R_X$ at the physical point, so we fix  the pion and nucleon mass to their physical values. We take the simple values $M_{\pi}=140$ MeV and $M_N=940$ MeV, unless stated otherwise. For the finite spatial volume we assume two lattice sizes such that $M_{\pi}L=4$ and $M_{\pi}L=6$.  The larger value is motivated by the simulation setup of the PACS-CS collaboration \cite{Ishikawa:2015rho}.

The ratio \pref{simplnotationR}  also depends on $n_{\rm max}$, the upper limit for the number of $N\pi$ states taken into account in the sum.  $n_{\rm max}$ should be chosen large enough such that the contribution from the omitted states is small and can be ignored. 
This essentially requires the times $t$ and $t'$ to be large enough such that the contribution of the omitted states is sufficiently suppressed in the ratio  $R_X$. 

ChPT puts an additional constraint on $n_{\rm max}$. Finite volume ChPT is an expansion in $p_n/\Lambda_{\chi}$, where the chiral scale $\Lambda_{\chi}$ is typically identified with $4\pi f_{\pi}$ \cite{Colangelo:2003hf}.
Thus, $n_{\rm max}$ is also constrained by insisting on a sufficiently small value for $p_{n_{\rm max}}/4\pi f_{\pi}$. 
In Ref.\ \cite{Bar:2015zwa} the condition $p_{n_{\rm max}}/\Lambda_{\chi}= 0.3$ was imposed for a reasonably well behaved chiral expansion. This bound translates into $n_{\rm max} =2$ and 5 for $M_{\pi}L=4$ and 6, respectively. Another reason for this particular bound is that the energy $E_{N\pi,{n_{\rm max}}}$ of the $N\pi$ states satisfying it is sufficiently well separated from the energy of the first resonance state, which is approximately 1.5$M_N$. In that case we can ignore mixing effects with the resonance state that is not contained in the chiral effective theory. 

There is some arbitrariness in imposing a bound on the momenta and the resulting values for $n_{\rm max}$.
Following Ref.\ \cite{Bar:2016uoj} we consider two additional values $n_{\rm max}$, specified in table \ref{table:nmax}. The largest one corresponds to $p_{n_{\rm max} }/ \Lambda_{\chi}\approx 0.6$. This is certainly not a small number and we do not expect a well-behaved chiral expansion in that case. Still, it turns out that for source-sink separations between 1 and 2 fm one essentially needs to include that many $N\pi$ states to saturate the sum in \pref{simplnotationR}. 
Note also that the two larger values $n_{\rm max}$  imply energies $E_{N\pi,{n_{\rm max}}}$ above the energy of the first resonance. Including $N\pi$ states with such high energies without taking into account the effect of the resonance is an approximation, and the results derived from it need to be interpreted with care.

%%% Table 
\begin{table}[tp]
\begin{center}
\begin{tabular}{l|c|c|c}
\multirow{2}{*}{$\frac{p_{n_{\rm max}}}{\Lambda_{\chi}}$}& \multicolumn{2}{|c|}{$n_{\rm max}$ } &  \multirow{2}{*}{$\frac{E_{N\pi,{n_{\rm max}}}}{M_N}$} \\ 
& $M_{\pi}L=4$ & $M_{\pi}L=6$ & \\  \hline
0.3 & 2 & 5 & $\approx 1.35 $ \\
0.45 & 5 & 12 & $\approx 1.6\phantom{3}$\\
0.6 & 10 & 22 & $\approx 1.9\phantom{3}$
\end{tabular}
\end{center}
\caption{$n_{\rm max}$ and $E_{N\pi,{n_{\rm max}}}$ as a function of $p_{n_{\rm max}}/\Lambda_{\chi}$, see main text.}
\label{table:nmax}
\end{table}
%%%

In practice there are two widely used methods to extract the moments, the plateau and the summation method. Both methods are based on the ratio $R_X$ as input. In the following we will consider only the plateau method. Applying ChPT and our results to the summation method requires very large source-sink separations, much larger than currently accessible in lattice QCD simulations (see section \ref{CommentSumMethod}).

\subsection{Impact on the plateau method}
For  a given source-sink separation $t$ the $N\pi$-state contribution to $R_X$ is minimal if the operator insertion time $t'$ is in the middle between source and sink. Therefore, the best estimate for the moments is the `midpoint' value $R_X(t,t/2)$. This midpoint estimate is essentially equivalent to what is called the `plateau estimate' in lattice determinations, and we will use this terminology here as well. 

\begin{figure}[p]
\begin{center}
\mbox{$R_X/\Pi_X$}\\
\includegraphics[scale=0.9]{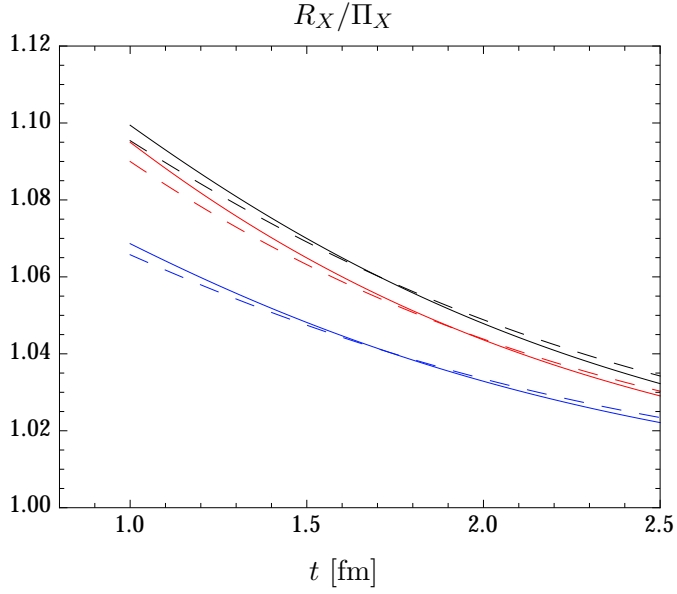}\\
\mbox{$t$ [fm]}
\caption{The plateau estimate $R_X(t,t/2)$ normalized by $\Pi_X$ for all three charges ($X=V$ in black, $A$ in blue, $T$ in red). Results for $M_{\pi}=140$ MeV and for $M_{\pi}L=4$ (solid lines) and $M_{\pi}L=6$ (dashed lines). $n_{\rm max}$ according to the first row in table \ref{table:nmax}.}
\label{fig:midpoint140}
\end{center}
\end{figure}

Figure \ref{fig:midpoint140} shows $R_X(t,t/2)/\Pi_X$, the plateau method estimate divided by the asymptotic value \pref{RAsympt} proportional to the moment. Without the $N\pi$ contribution this ratio is equal to 1. Any deviation from 1 is the relative error caused by the $N\pi$-state contribution. Plotted are the results for all three moments ($X=V,A,T$) for $M_{\pi}L=4$ and $M_{\pi}L=6$. We can make the following observations: (i) The differences between the results for the two different volumes are rather small, much smaller than the expected accuracy of the LO results. 
(ii) All three curves are above 1, so the plateau estimates overestimate the moments in all three cases. 

% Figure
% End figure

% Figure
\begin{figure}[p]
\begin{center}
\mbox{$R_V/\Pi_V$}\\
\includegraphics[scale=0.9]{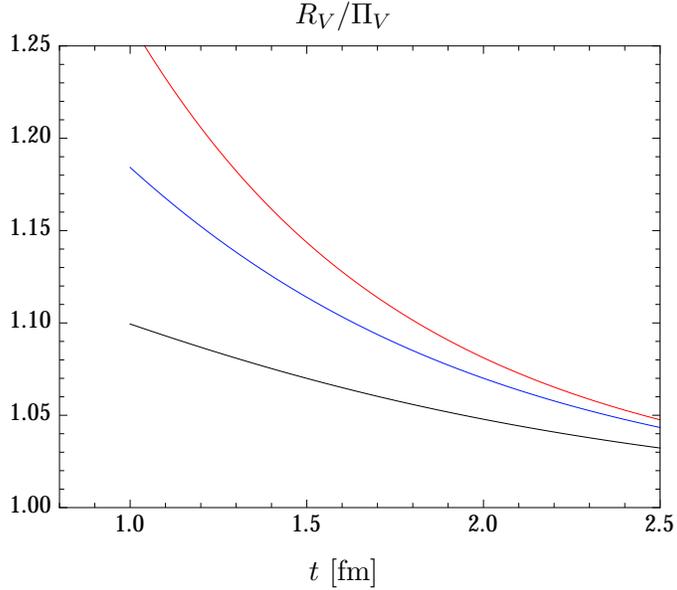}\\
\mbox{$t$ [fm]}
\caption{The plateau estimate $R_V(t,t/2)$ normalized by $\Pi_V$ for $M_{\pi}=140$ MeV, $M_{\pi}L=4$ and the three different $n_{\rm max}$ values specified in table \ref{table:nmax} ($n_{\rm max}=2$ in black, 5 in blue and 10 in red).}
\label{fig:midpointV3nmax}
\end{center}
\end{figure}
% End figure

Qualitatively the same results have been found for the nucleon charges  \cite{Bar:2016uoj}. There too the $N\pi$ contribution leads to an overestimation of the charges, and the finite-volume dependence was found to be equally small. As already discussed in  \cite{Bar:2016uoj} the  small dependence on the volume requires that the energy interval $[M_N+M_{\pi}, E_{{N\pi},n_{\rm max}}]$ of the $N\pi$ states taken into account for $R_X$ is kept constant as a function of the volume. This is the case for the $n_{\rm max}$ values in table \ref{table:nmax}.  

Figure \ref{fig:midpoint140} shows that the results for the vector and tensor operator are very close and about 50\% larger than the result for the axial vector operator. In contrast, the HB limit predicts the same $N\pi$ contribution for $X=A$ and $T$, cf.\ \pref{HBlimitValues} . The corrections to the HB limit make both $R_V$ and $R_A$ smaller and $R_T$ larger, leading to the curves in  fig.\ \ref{fig:midpoint140}. 

Figure \ref{fig:midpointV3nmax} shows the impact of the $N\pi$ states as $n_{\rm max}$ is increased in case of the vector operator, i.e. for the determination of the momentum fraction $\mf$.  Results are shown for the three $n_{\rm max}$ values specified in table \ref{table:nmax} for $M_{\pi}L=4$. The analogous results  for $M_{\pi}L=6$ lie essentially on top of the curves in fig.\ \ref{fig:midpointV3nmax} and are not displayed.  
To a good approximation $n_{\rm max}=10$ (red curve) saturates the sum in the ratio; adding more states does not change the result significantly, at least for the sink times considered in the plot. Therefore, we call this the full $N\pi$ contribution for short.

Figure \ref{fig:midpointV3nmax} shows clearly what we remarked before: The larger $t$ the smaller the impact of the high momentum $N\pi$ states relative to the impact of the lowest two states. At $t=2$ fm the contribution from the first two states (black curve) makes approximately 60\% of the full contribution (red curve). This ratio increases to about 70\% at $t=2.5$ fm. At source-sink separations as large as this we may ignore all but the lowest two states. For those we expect our LO result to give a reasonable estimate for the $N\pi$ contribution; the NLO correction is O($p^2$) and one may expect, as a rough estimate, a 30\% correction. A more honest error estimate requires the result of the NLO calculation.

% Figure
\begin{figure}[p]
\begin{center}
\mbox{$R_A/\Pi_A$}\\
\includegraphics[scale=0.9]{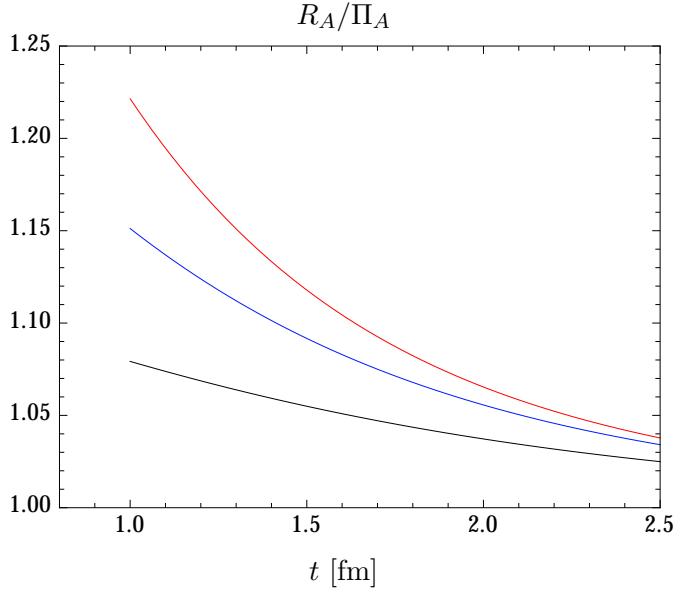}\\
\mbox{$t$ [fm]}
\caption{The plateau estimate $R_A(t,t/2)$ normalized by $\Pi_A$ for $M_{\pi}=140$ MeV, $M_{\pi}L=4$ and the three different $n_{\rm max}$ values specified in table \ref{table:nmax} ($n_{\rm max}=2$ in black, 5 in blue and 10 in red).}
\label{fig:midpointA3nmax}
\end{center}
\end{figure}
% End figure

% Figure
\begin{figure}[p]
\begin{center}
\mbox{$R_T/\Pi_T$}\\
\includegraphics[scale=0.9]{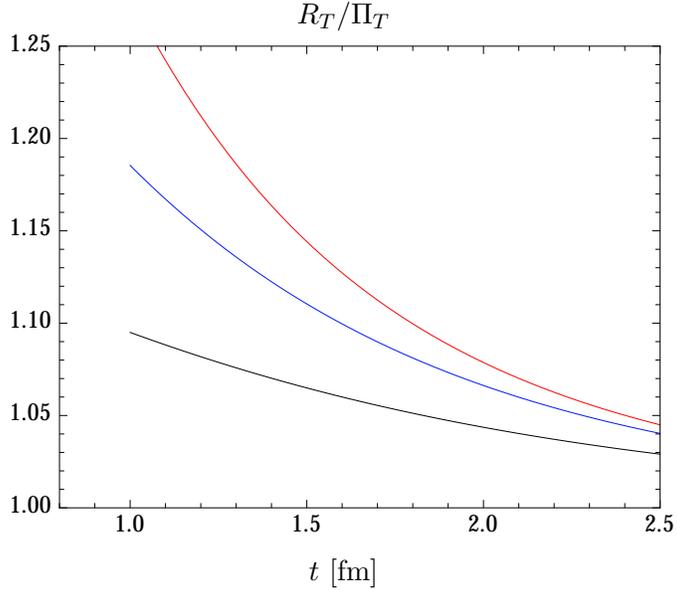}\\
\mbox{$t$ [fm]}
\caption{The plateau estimates $R_T(t,t/2)$ normalized by $\Pi_T$ for $M_{\pi}=140$ MeV, $M_{\pi}L=4$ and the three different $n_{\rm max}$ values specified in table \ref{table:nmax} ($n_{\rm max}=2$ in black, 5 in blue and 10 in red).}
\label{fig:midpointT3nmax}
\end{center}
\end{figure}
% End figure

For $t$ less than 2 fm the impact of the higher momentum $N\pi$ states increases rapidly. At $t=1.5$ fm the lowest two states contribute less than 50\% to the full contribution. Since the contribution of the high-momentum $N\pi$ states is prone to larger NLO corrections we can only give a crude estimate. Reading off a +15\%  $N\pi$ contribution at $t=1.5$ fm and allowing for a 50\% error due to higher order corrections we would arrive at a 10-20\% overestimation of $\mf$ at $t$ about $1.5$ fm. As before, the error estimate of 50\% is a naive guess that can be put on firmer grounds with a calculation at NLO.

For $t$ smaller than 1.5 fm the higher momentum $N\pi$ states rapidly dominate the $N\pi$ contribution and we do not expect our LO ChPT result to be a reasonable approximation anymore. It is also likely that working to higher order in the chiral expansion will not help in going to such small source-sink separations. However, we may still conclude that as many as 10 $N\pi$ states contribute substantially to the ratio $R_V$ for source-sink separations between 1 and 1.5 fm, a slightly unsettling high number.

% Figure
\begin{figure}[p]
\begin{center}
\mbox{$R_V/\Pi_V$}\\
\includegraphics[scale=0.9]{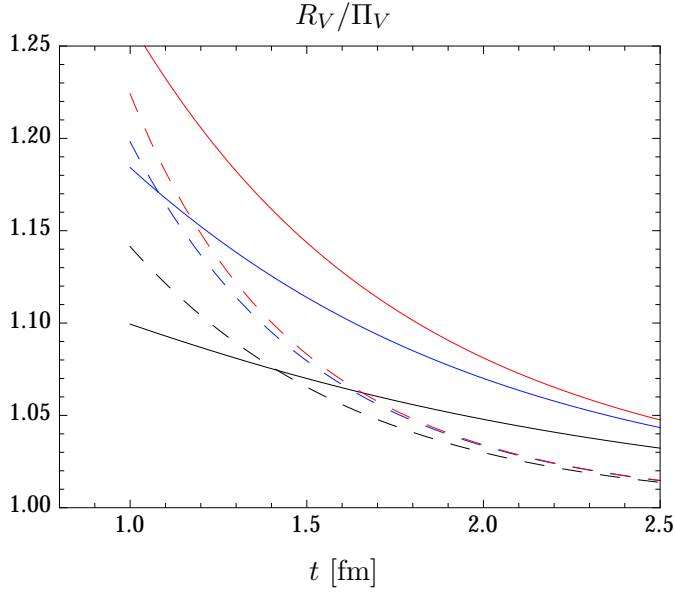}\\
\mbox{$t$ [fm]}
\caption{The plateau estimates $R_V(t,t/2)$ normalized by $\Pi_V$ for $M_{\pi}/M_N=0.149$ (solid lines) and $M_{\pi}/M_N=0.27$ (dashed lines). In both cases $M_{\pi}L=4$, and the three different $n_{\rm max}$ values specified in table \ref{table:nmax} are used ($n_{\rm max}=2$ in black, 5 in blue and 10 in red).}
\label{fig:midpointV3nmax2masses}
\end{center}
\end{figure}
% End figure

% Figure
\begin{figure}[p]
\begin{center}
\mbox{$R_X/\Pi_X$ and $R_X/g_X$}\\
\includegraphics[scale=0.9]{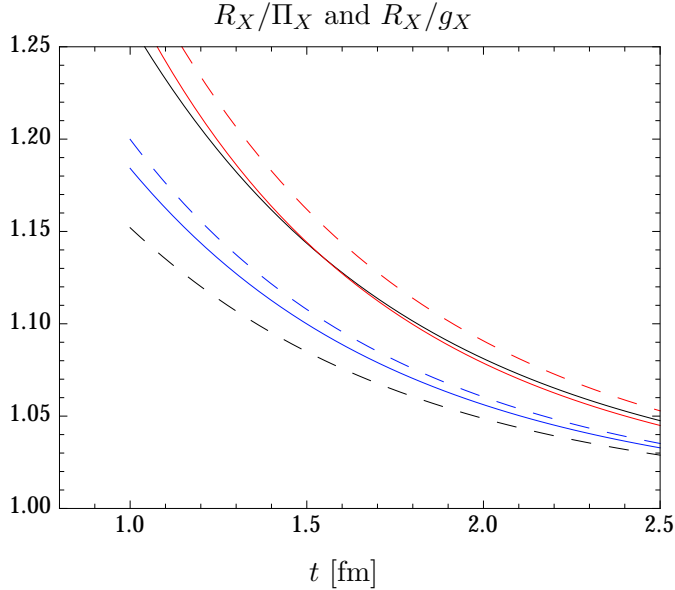}\\
\mbox{$t$ [fm]}
\caption{The ratios $R_X(t,t/2)/\Pi_X$ for the three moments (solid lines, $X=V$ in black, $A$ in blue, $T$ in red; same curves as in fig.\ \ref{fig:midpoint140}) compared to the ratios $R_X(t,t/2)/g_X$ for the three nucleon charges (dashed lines, $X=A$ in black, $T$ in blue, $S$ in red). Results for $M_{\pi}=140$ MeV and for $M_{\pi}L=4$ with $n_{\rm max} =10$.}
\label{fig:compare}
\end{center}
\end{figure}
% End figure

The results for  the helicity and transversity moment are shown in figs.\ \ref{fig:midpointA3nmax} an \ref{fig:midpointT3nmax}, respectively. In case of the transversity moment $\tm$ the differences to the momentum fraction are only marginal. The results for the helicity moment are about 30\% smaller. Therefore, following the reasoning given before in case of $\mf$ we would arrive at a 7-15 \% overestimation of $\hm$ due to the $N\pi$ contribution at about $1.5$ fm.

Since we are mainly interested in lattice simulations with physical pion masses we kept the ratio $M_{\pi}/M_N$ fixed at its approximate physical value 0.149. For a larger than physical pion mass one expects the $N\pi$ contribution to become rapidly smaller. As an illustration for this behavior fig.\ \ref{fig:midpointV3nmax2masses} shows again the results for the momentum fraction 
at the physical point (solid lines) compared to those 
for $M_{\pi}/M_N =0.27$ (dashed lines), a value close to the one found by the 
RQCD collaboration in their simulations with a pion mass of about 295 MeV.\footnote{See table 1 in Ref.\ \cite{Bali:2014nma}, results for ensemble IV.}   Since the pion mass is larger the energies $E_{N\pi,n}$ of the $N\pi$ states are larger than in the case with a physical pion mass. If we keep $M_{\pi}L=4$ fixed also the spatial volume is smaller, implying larger discrete spatial momenta of the moving nucleon and pion. Therefore, the energy gaps $\Delta E_n$ in \pref{simplnotationR} are larger and a faster exponential suppression of the $N\pi$ contribution is expected.

Figure \ref{fig:midpointV3nmax2masses} supports this expectation, at least for $t$ not smaller than 1.5 fm. The full $N\pi$ contribution is significantly reduced; at $t=2.5$ fm by a factor of four (red curves). Moreover, the impact of the higher momentum $N\pi$ states is drastically reduced. Even at $t=1.5$ fm the contribution from the first two states (dashed black curve) makes 80\% of the full contribution (dashed red curve). However, for smaller $t$ the contribution of the higher momentum states increases rapidly again. Interestingly, the curves for the contribution of the lowest two states (black curves) cross at $t\approx 1.4$ fm. So despite the larger energy gaps for the heavier pion mass the lowest two states have a larger impact than the lowest two states for the physical pion mass. The reason are the larger values for the coefficients $b_{V,n}$ for a heavier pion mass, which are here about a factor   
2.5 to 3 larger than their analogues for the physical pion mass. Even though we need to be careful with drawing conclusions from our LO results at such small $t$ values, this example serves as a warning that prejudices about excited-state contributions based on the energy gaps alone can be quite misleading.   

Finally, figure \ref{fig:compare} compares the results for the moments with the analogous ones for the nucleon charges \cite{Bar:2016uoj}. Results are shown for $M_{\pi}L=4$ only. As already mentioned, the $N\pi$-contribution results in an overestimation for all moments and charges. Qualitatively the observables can be separated in two groups. The $N\pi$-contribution for the scalar charge,  the momentum fraction and the transversity moment are larger by about 50\% compared to the contribution in the axial and tensor charge and the transversity moment. The $N\pi$-contribution is smallest for the axial charge and largest for the scalar charge.

\subsection{Comment on the summation method}\label{CommentSumMethod}

The summation method \cite{Maiani:1987by,Capitani:2012gj} starts from the ratio $R_X$ and computes the integral $S_X(t,t_{\rm m}) = \int^{t-t_{\rm m}}_{t_{\rm m}} dt' R_X(t,t')$.  As a function of $t$ (keeping $t_{\rm m}$ fixed) the slope is proportional to the moment one is interested in.
In actual lattice determinations $t_{\rm m}$ is taken to be zero, so the integral is computed for all insertion times $t'$ between source and sink. On the other hand, for ChPT to give a good approximation of $R_X$ all time separations need to be large. Based on the results in the last section we need to require a minimal time separation of about 1 fm for $t_{\rm m}$ and $t-t_{\rm m}$. In addition we need a non-zero time interval $t-2t_{\rm m}$ to integrate over. This implies source-sink separations of at least 2.5 fm if not larger. Such large values are currently not accessible in lattice simulations, so at present it would be purely academic to study the $N\pi$-state contribution to the summation method.

\section{Conclusions}
As already mentioned in the introduction, some collaborations have already performed lattice calculations of the various moments with physical or near-to-physical pion masses \cite{Green:2012ud,Bali:2014gha,Abdel-Rehim:2015owa}.  The main obstacle for applying the results found here to these calculations are the fairly small source-sink separations $t$ in these simulations. The maximal source-sink separation $t_{\rm max}$ used to extract the moments with the plateau method ranges between 1.1 fm and 1.3 fm. As discussed in the previous section, we do not expect LO ChPT to provide solid results for source-sink separations that small. Still, it is worth to emphasize a few observations.

The lattice results of the quark momentum fraction and the helicity moment are typically larger than their phenomenological values.
In case of $\mf$ the overestimation is about 20-30\%, in case of the helicity moment it is somewhat smaller. Thus, qualitatively the overestimation due to the $N\pi$ states goes into the right direction. For source-sink separations of about 1.5 fm we estimated the overestimation to 10-20\%  in case of $\mf$. Even though not very precise this estimate suggests that the $N\pi$-state contribution may form a substantial part of the total excited-state contamination presently observed in lattice data. 

Compared with the results for the nucleon charges we find the $N\pi$ contribution to the scalar charge $g_S$ to be the largest one.
Also this is qualitatively in agreement with what is observed in lattice calculations, for example in Ref.\ \cite{Abdel-Rehim:2015owa}. 
However, one should also emphasize that LO ChPT predicts an overestimation of the axial charge $g_A$ due to $N\pi$ states, in conflict with the lattice estimates that typically underestimate the experimental value.\footnote{A recent attempt to explain this apparent contradiction can be found in Ref. \cite{Hansen:2016qoz}.}  This serves once again as a warning that the source-sink separations realized in present-day lattice simulations are probably too small for the LO ChPT results to apply.

Larger source-sink separations in lattice simulations obviously help in reducing the impact of excited states and in making contact with the ChPT results derived here. Continuous progress is being made with lattice measurements at larger source-sink separations. The ETM collaboration, for example, has recently announced lattice results for the moments at $t\approx 1.7$ fm \cite{Alexandrou:2016hiy}. 
Still, the excited-state suppression may not be as efficient as one is hoping for. Taking once again $\mf$ as an example we still expect an overestimation of about 10\% for source-sink separations of approximately 2 fm. Such large time separations seem out of reach with current lattice techniques, and new methods to increase the signal-to noise ratio in lattice simulations are needed. A recent proposal \cite{Ce:2016idq,Ce:2016ajy} to factorize the fermion determinant and propagator in lattice QCD together with multilevel Monte Carlo integration methods seems very promising in that respect. 
  
The results derived here are based on LO ChPT. Working out the NLO correction is certainly desirable because it will provide stronger error estimates for the LO results. In addition, the impact of other multi-hadron states ($\Delta\pi$, $N\pi\pi$) and the Roper resonance need to be studied as well. The chiral effective theories including the $\Delta$ and the Roper are known and the calculations will be analogous to the one presented here. Once all these contributions are taken into account one can expect ChPT to provide reliable estimates for the excited-state contaminations due to multi-particle states, that, hopefully, can also be used to analytically remove them from the lattice results.     

\vspace{2ex}
\noindent {\bf Acknowledgments}
\vspace{2ex}

Correspondence with Peter Bruns is gratefully acknowledged. I also thank the Yukawa Institute for Theoretical Physics for its kind hospitality. This work is supported by the Japan Society for the Promotion of Science (JSPS) with an Invitation Fellowship for Research in Japan (ID No. L16520).
\vspace{3ex}
 
\begin{appendix}
\section{The tensor operator in Baryon ChPT}\label{appTensor}
 
In the following we outline the mapping of the QCD tensor operator in eq.\ \pref{tensorop} onto its ChPT analogue (to LO) in \pref{OpTensor}.
The mapping follows the general procedure: We first introduce a source term for the tensor operator that is added to the massless QCD lagrangian. Subsequently, this source term is mapped to ChPT taking into account its transformation properties under chiral symmetry, parity and charge conjugation. 

In terms of chiral quark fields the source term reads 
\begin{equation}\label{Lexttensor}
{\cal L}_{\rm tensor}=\overline{\psi}_R t^{RL}_{\mu\nu\rho}\sigma_{{[\mu\{\nu]}}D^-_{\rho\}} \psi_L +  \overline{\psi}_L t^{LR}_{\mu\nu\rho}\sigma_{{[\mu\{\nu]}}D^-_{\rho\}}\psi_R 
\end{equation}
with two matrix valued source fields $t^{RL}_{\mu\nu\rho}= t^{RL,a}_{\mu\nu\rho}T^a$ and $t^{LR}_{\mu\nu\rho}= t^{LR,a}_{\mu\nu\rho}T^a$. They couple left- and right handed fields as indicated by the superscripts. The tensor operator \pref{tensorop} is obtained from the source term by taking derivatives with respect to the two source fields and adding the two contributions.

The symmetrization and antisymmetrization that is associated with the curly and square brackets in the operator can be transferred to the source field, i.e.\ $t^{RL}_{\mu\nu\rho}\sigma_{_{[\mu\{\nu]}}D^-_{\rho\}} = t^{RL}_{[\mu\{\nu]\rho]}\sigma_{\mu\nu}D^-_{\rho} $. In order to keep the notation simple we drop the curly and square brackets in the following but keep in mind the symmetry properties of the source fields. 

Under chiral transformations $R,L$ the source term is invariant if the source fields transform according to  
$t^{RL}_{\mu\nu\rho} \longrightarrow Rt^{RL}_{\mu\nu\rho}L^{\dagger}$ and  $t^{LR}_{\mu\nu\rho} \longrightarrow L t^{LR}_{\mu\nu\rho} R^{\dagger}$. In addition, the source term is invariant under parity ($P$) and charge conjugation ($C$) provided the source fields transform according to $t^{RL}_{\mu\nu\rho} \longrightarrow t^{LR}_{\mu\nu\rho}$ under $P$ and $t^{RL}_{\mu\nu\rho} \longrightarrow [t^{LR}_{\mu\nu\rho}]^T$ under $C$, where $T$ refers to taking the transpose in flavor space. 

Based on these symmetry properties the source term can be mapped to ChPT. 
It is useful to introduce the combinations
\begin{equation}\label{Deftmunupm}
t^{\pm}_{\mu\nu\rho} = u^{\dagger} t^{RL}_{\mu\nu\rho}u^{\dagger} \pm u t^{LR}_{\mu\nu\rho}u\,,
\end{equation}
with $u$ being the standard chiral field involving the pion fields. The reason for this definition is that the fields $t^{\pm}_{\mu\nu,\rho}$ transform as $t^{\pm}_{\mu\nu\rho} \longrightarrow h t^{\pm}_{\mu\nu\rho} h^{-1}$ under chiral transformations, where $h$ denotes the compensator field associated with the non-linear realization of chiral symmetry \cite{Coleman:1969sm,Callan:1969sn}. Under $P$ and $C$ the source fields in \pref{Deftmunupm} transform as the original source fields. 

Invariants under chiral symmetry are now easily constructed. We find it convenient to follow Ref.\ \cite{Fettes:2000gb}.\footnote{Ref.\ \cite{Fettes:2000gb} assumes the Minkowski space-time metric. For the main construction principle this is irrelevant and we transcribe the necessary formulae to the euclidean space-time metric. Except for this modification we follow the conventions and notation in Ref.\ \cite{Fettes:2000gb}.} According to section 2.2.\ of that reference any invariant monomial in the effective $N\pi$ Lagrangian is of the generic form
\begin{equation}
\overline{\psi}A_{\mu\nu\ldots} \Theta_{\mu\nu\ldots} \psi + {\rm h.c.}\,\,.
\end{equation}
Here $A_{\mu\nu\ldots}$ is a product of pion and/or source fields and their covariant derivatives. $\Theta_{\mu\nu\ldots}$ is a product of a Clifford algebra elements and a totally symmetrized product of covariant derivatives acting on the nucleon fields. These building blocks obey various restrictions stemming from chiral symmetry. In addition, the equations of motion can be used to remove terms in the chiral lagrangian that are redundant. 

Here we are interested in the leading terms involving the tensor source field only once. The simplest terms with lowest chiral dimension are obtained with $A_{\mu\nu\rho} = t^{+}_{\mu\nu\rho}$. A list of independent rank 3 tensor structures $\Theta_{\mu\nu\rho}$ is given in eq.\ (A.21) of Ref.\ \cite{Fettes:2000gb},
\begin{equation}\label{list}
\delta_{\mu\nu}\gamma_5\gamma_{\rho}, \delta_{\mu\nu}D_{\rho}, \sigma_{\mu\nu}D_{\rho}, \epsilon_{\mu\nu\rho\lambda} D_{\lambda},\gamma_5\gamma_{\mu}D_{\nu\rho}, D_{\mu\nu\rho}\,.
\end{equation}
The first two, the fourth and the last structure vanish once they are contracted with $t^{+}_{\mu\nu\rho}$ due to the symmetry properties of the source field. Making use of the equations of motion the fifth structure is equivalent to $\gamma_5\sigma_{\mu\nu} D_{\rho}$, see eq.\ (2.33) in Ref.\ \cite{Fettes:2000gb}. Since $\gamma_5\sigma_{\mu\nu} = \epsilon_{\mu\nu\alpha\beta}\sigma_{\alpha\beta}/2$ this structure is not independent of the third entry in the list \pref{list}. So we conclude that there is only one independent structure $\Theta_{\mu\nu\rho}=\sigma_{\mu\nu}D_{\rho}$, and this leads to the operator given in \pref{OpTensor} in section \ref{chpt}.

As already mentioned source terms involving pion fields only are necessarily beyond LO. The reason is the Lorentz indices can be provided only by partial derivatives of the pion fields.
 
 \section{Summary of Feynman rules}\label{appFeynmanRules}
 We employ the covariant formulation of baryon ChPT \cite{Gasser:1987rb,Becher:1999he}, and our calculations are done to LO in the chiral expansion. To that order the chiral effective lagrangian consists of two parts only, ${\cal L}_{\rm eff}={\cal L}_{N\pi}^{(1)} + {\cal L}_{\pi\pi}^{(2)}$. Expanding this lagrangian in powers of pion fields and keeping interaction terms with one pion field only we obtain
\begin{eqnarray}
\label{Leff}
{\cal L}_{\rm eff} &=& \overline{\Psi} \Big(\gamma_{\mu}\partial_{\mu} +\mN \Big)\Psi +\frac{1}{2}\pi^a \Big(- \partial_{\mu}\partial_{\mu} + M_{\pi}^2 \Big)\pi^a + \frac{ig_A}{2f}\overline{\Psi}\gamma_{\mu}\gamma_5\sigma^a \Psi \, \partial_{\mu} \pi^a\,.
\end{eqnarray}
The  nucleon fields $\Psi=(p,n)^T$ and $\overline{\Psi}=(\overline{p},\overline{n})$ 
contain the proton and the neutron fields $p$ and $n$.  $M_{\pi}$ denotes the pion mass, while $M_N$, $g_A$ and $f$ are the chiral limit values of the nucleon mass, the axial charge and the pion decay constant. 

The interaction term in \pref{Leff} leads to the well known nucleon-pion interaction vertex proportional to the axial charge. A factor $i$ appears here because we work in euclidean space-time. From the terms quadratic in the fields one reads off the nucleon and pion propagators. We find the time-momentum representation for the propagators convenient. In that representation the pion propagator reads
\begin{eqnarray}
G^{ab}(x,y)& = &  \delta^{ab}L^{-3}\sum_{\vec{p}} \frac{1}{2 E_{\pi}} e^{i\vec{p}(\vec{x}-\vec{y})} e^{-E_{\pi} |x_0 - y_0|}\,,\label{scalprop}
\end{eqnarray}
with  the pion energy given by $E_{\pi} =\sqrt{\vec{p}^2 +M_{\pi}^2}$. The nucleon propagator $S^{ab}_{\alpha\beta}(x,y)$ is given by
\begin{eqnarray}
S_{\alpha\beta}^{ab}(x,y)& = &  \delta^{ab} L^{-3}\sum_{\vec{p}} \frac{Z_{p,\alpha\beta}^{\pm}}{2E_N} e^{i\vec{p}(\vec{x}-\vec{y})} e^{-E_N |x_0 - y_0|}\,.
\end{eqnarray} 
$a,b$ and $\alpha,\beta$ refer to the isospin and Dirac indices, respectively. The factor $Z^{\pm}_{\vec{p}}$ in the nucleon propagator (spinor indices suppressed) is defined as
\begin{equation}
Z_{\vec{p}}^{\pm}=-i\vec{p}\cdot\vec{\gamma} \pm E_N \gamma_0+\mN\,, 
\end{equation}
where the $+$ ($-$) sign applies to $x_0 > y_0$ ($x_0 < y_0$), and the nucleon energy is given by $E_N=\sqrt{\vec{p}^2 +M_N^2}$. 
The sum in both propagators runs over the discrete spatial momenta that are compatible with periodic boundary conditions imposed on the finite spatial volume, i.e.\ 
$\vp=2\p\vn/L$  with $\vn$ having integer-valued components.

The expressions for the nucleon interpolating fields in ChPT have been derived in Ref.\ \cite{Wein:2011ix}. To LO and up to one power in pion fields one finds
\begin{eqnarray}\label{Neffexp}
N(x)& = & \tilde{\alpha} \left(\Psi(x) + \frac{i}{2f} \pi^a(x)\sigma^a \gamma_5\Psi(x)\right)\,,\\
\overline{N}(0) & = & \tilde{\beta}^* \left(\overline{\Psi}(0) + \frac{i}{2f}\overline{\Psi}(0)\gamma_5\sigma^a\pi^a(0) \right)
\end{eqnarray}
These are the effective fields for the standard nucleon interpolating fields composed of three quarks without derivatives \cite{Ioffe:1981kw,Espriu:1983hu}. The interpolating fields not necessarily need to be point-like, but can also be constructed from `smeared' quark fields. These operators map to the same chiral expressions provided the smearing procedure is compatible with chiral symmetry and the `smearing radius' is small compared to the Compton wavelength of the pion. In that case smeared interpolating fields are mapped onto point like fields in ChPT just like their pointlike counterparts at the quark level  \cite{Bar:2013ora,Bar:2015zwa}. The expressions differ only by the LECs $\tilde{\alpha},\tilde{\beta}$. If the same interpolating fields are used at both source and sink we find $\tilde{\alpha}=\tilde{\beta}$. 

\end{appendix}

\end{document}